# Generalised Framework for Controlling and Understanding Ion Dynamics with Passivated Lead Halide Perovskites

Tomi K. Baikie[1], Philip Calado[2], Krzysztof Galkowski[1,3,4], Zahra Andaji-Garmaroudi[1], Yi-Chun Chin[2], Joel Luke[2], Charlie Henderson[2], Tom Dunlop[5], James McGettrick[5], Ji-Seon Kim[2], Akshay Rao[1], Jenny Nelson[2], Samuel D. Stranks[1,6], Piers R. B. Barnes[2]

1 - Cavendish Laboratory, University of Cambridge, J. J. Thomson Avenue, Cambridge, CB3 0HE, UK

2 - Department of Physics, Imperial College London, London, SW7 2AZ, UK

3 - Department of Experimental Physics, Faculty of Fundamental Problems of Technology, Wroclaw University of Science and Technology, Wroclaw, Poland

4 - Institute of Physics, Faculty of Physics, Astronomy and Informatics, Nicolaus Copernicus University, Toruń, Poland

5 - SPECIFIC IKC, Faculty of Science and Engineering, Swansea University, Fabian way, Swansea, SA1 8EN, UK.

6 - Department of Chemical Engineering and Biotechnology, University of Cambridge, Cambridge, UK

Abstract:

Metal halide perovskite solar cells have gained widespread attention due to their high efficiency and high defect tolerance. The absorbing perovskite layer is as a mixed electron-ion conductor that supports high rates of ion and charge transport at room temperature, but the migration of mobile defects can lead to degradation pathways. We combine experimental observations and drift-diffusion modelling to demonstrate a new framework to interpret surface photovoltage (SPV) measurements in perovskite systems and mixed electronic ionic conductors more generally. We conclude that the SPV in mixed electronic ionic conductors can be understood in terms of the change in electric potential at the surface associated with changes in the net charge within the semiconductor system. We show that by modifying the interfaces of perovskite bilayers, we may control defect migration behaviour throughout the perovskite bulk. Our new framework for SPV has broad implications for developing strategies to improve the stability of perovskite devices by controlling defect accumulation at interfaces. More generally, in mixed electronic conductors our framework provides new insights into the behaviour of mobile defects and their interaction with photoinduced charges, which are foundational to physical mechanisms in memristivity, logic, impedance, sensors and energy storage.

Introduction:

In applications of mixed ionic electronic conductors, such as metal halide perovskite solar cells, understanding and controlling the movement of both electrical and ionic charge is crucial. Metal halide perovskite solar cells are promising materials with high defect tolerance, but effective management of ions are not only important for understanding degradation processes in solar cells and LEDs [1]–[4],but also for their proper operation as X-ray detectors[5] and logic devices[6], [7]. The perovskite absorbing layer in these devices exhibits a wide range of defects and, as a portion of these charged defects are mobile, these layers can support high rates of ion and charge transport at room temperature [8]–[10]. Ion migration underpins a wide range of reported observations in perovskite devices, including conductivity [8], [11], [12], atmospheric stability [13], hysteresis [8], [14], [15], and functionality in synaptic networks [16] and memristors [8]. Notably, recent work has shown that mobile defects may induce degradation, which significantly inhibit device performance (see **Figure 1**) [17]–[20]. In doped semiconductors, surface photovoltage (SPV) measurements have been extensively used to monitor the behaviour of charge carriers at interfaces. Consequently, this technique appears attractive in the context of perovskites, but is not yet well understood in the context of mixed ion-electronic conductors.

Using a new framework to interpret surface photovoltage (SPV) measurements in semiconductor systems with buried junctions and mixed conducting behaviour, we demonstrate it is possible to control defect migration behaviour through the perovskite bulk by modifying the interfaces. We illustrate the mechanism by changing the interface properties of thin film bilayers and photovoltaic devices. Our observations and explanation have broad implications both for the interpretation of the SPV technique when applied to mixed conducting semiconductors as well as for developing strategies to control defect accumulation at interfaces in working perovskite devices – a critical concern for improving their stability. The interpretation of SPV presented here ulocks a deeper understanding of device functionality, particularly those leveraging the attributes of mixed electronic and ionic conductivity and holds promise as a valuable tool for advancing the up and coming electronic devices.

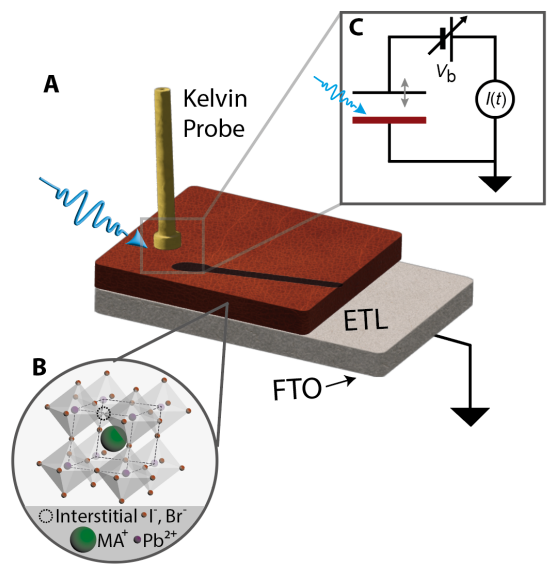

*Figure 1* – A Schematic of a halide perovskite (inset B) film deposited on an electron transport layer (ETL), upon a grounded electrode such as ITO or FTO. The tip of a Kelvin probe is shown close to the top surface. The electrical potential of the surface relative to the tip is measured before, during and after illumination, and the resulting change in potential during illumination gives the "surface photovoltage" signal. Inset C – simplified circuit diagram. The probe vibrates above the kelvin probe, inducing a current I(t). A backing voltage $V_b$ nulls the current, which gives the contact difference between the tip and the sample.

Both in air and vacuum, the SPV response for each condition studied shows a clear biphasic change with time over a few minutes (**Figure 2A&E**), and after tens of minutes tends towards steady-state. The nitrogen atmosphere results are presented in **SI Section 1**. Surprisingly, the polarity of the SPV signal was reversed by the atmosphere. When the illumination ended, the SPV signal decayed to the dark equilibrium level over several hours – on a longer timescale than the rise. These results are consistent between samples, although the magnitude of the signal varied from batch to batch (further details and replicate analysis is reported in **SI Section 4**). Concurrent photoluminescence (PL) measurements were also made to monitor changes in the relative fraction of radiative to non-radiative recombination in the samples during the SPV experiment (**SI Figure 4**). As previously reported, an increase in PL was seen during illumination under oxygen, consistent with passivation of non-radiative recombination sites, thus increasing the fraction of radiative recombination[21]. In contrast, a reduction in the PL signal relative to the initial state is seen during illumination under vacuum, indicating an increase in the number of non-radiative recombination centres and a reduction of electronic charge density in the perovskite bulk [21].

Our observations are consistent with previous reports from Daboczi *et al.* who contrasted the effect of a hole transport layer (HTL) with that of an electron transport layer (ETL) on the SPV measurement. Daboczi *et al.* observe an inversion in the SPV polarity when using PEDOT:PSS, an HTL, when compared to $TiO_2$, an ETL[22, Figure 3 & 5]. These results are also consistent with results from Harwell *et al.,* who also observe the characteristic biphasic response of perovskite bilayers deposited on TiO$_2$ [23, Figure 3]. Hu *et al.* reported the SPV response on ETL and HTLs, observing an inversion in the SPV response, as we have, when changing from ambient to vacuum conditions [24, Figure 4]. The dark response not returning as quickly to the original value is consistent with reports suggesting the formation of lead oxide under illumination[18], however since we also see an asymmetry in relaxation under vacuum (where oxide formation might be precluded) a process which is at least partially reversible and unrelated to oxygen is suggested in this case.

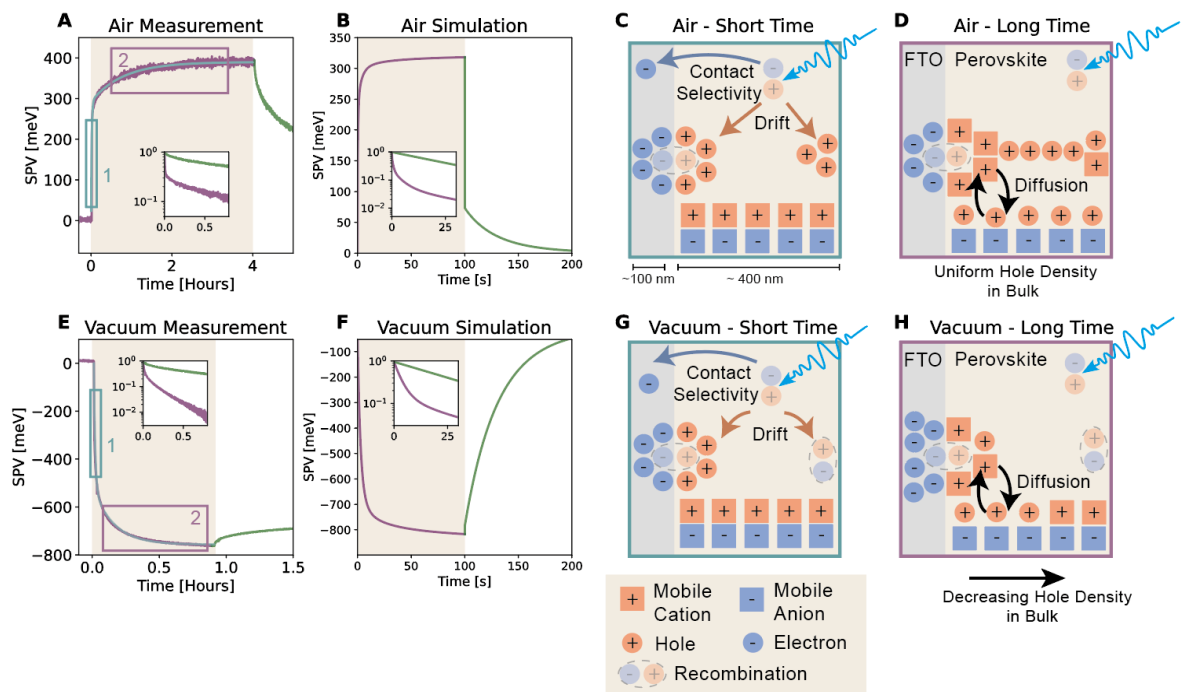

*Figure 2 Surface photovoltage measurements in air and under vacuum.* SPV measurements and simulations of a perovskite $FA_{0.79}MA_{0.15}Cs_{0.06}Pb(I_{0.85}Br_{0.15})_3$ /FTO bilayer under illumination (purple) and dark measurement (green). The beige overlays indicates the period when the sample was illuminated with approximately 60 mW/cm$^2$ of 532 nm laser light. The insets show the normalised absolute SPV signal plotted on a log scale for both the illuminated and dark components. The sample was measured in (**A**) air and (**B**) under vacuum. The corresponding simulations are shown where surface recombination velocity was either set to zero (i.e. surface passivated, **C**) or left at a standard value (i.e. the surface is not passivated, **D**) the perovskite/FTO interface recombination velocity was constant for both cases. **E – H** Schematics depicting the mechanisms underlying the SPV signal observed in the simulations. They represent the changes that occur relative to dark equilibrium and mobile ionic charge concentrations exceed electronic concentrations. The FTO acts as a selective contact to extract photogenerated electrons from the perovskite layer resulting in an accumulation of holes in the perovskite. A dynamic equilibrium is rapidly reached where the rate of recombination of electrons and holes approximately equals the rate of photogeneration. If the surface is passivated (**E**), the SPV signal is positive due to a net accumulation of holes at the interface and surface of the sample. At longer times (**G**) ions slowly diffuse to exchange with holes at the interface leading to a net increase accumulated of holes in the perovskite to maintain the same recombination rate with a corresponding increase in SPV. If the surface is unpassivated (**F**), the additional surface recombination pathway results in a lower concentration of photogenerated holes in the perovskite and consequently an accumulation of electrons in the FTO to maintain the same net recombination rate. This leads to a negative SPV. At longer times (**H**), positive mobile ions diffusively exchange with holes at the surface requiring further accumulation of photogenerated electrons in the FTO to maintain the net recombination rate leading to a more negative SPV.

The traditional interpretation of SPV signals does seem not capable of explaining the change in polarity observed in these observations. The standard interpretation focusses on changes in the surface space charge layer of the semiconductor under illumination (normally related to changes in the concentration of trapped surface charge). This explanation is inadequate for semiconductors which show mixed electronic-ionic conduction, and indeed semiconductor structures which have 'buried' space charge regions in addition to any surface charging (see discussion in **SI Section 6**). Here we propose a new conceptual framework underpinned by drift-diffusion simulations to understand SPV signals in samples including mixed conductors such as perovskites and layer structures with internal space charge regions.

The SPV signal can only be correlated to a change in the electrostatic potential at the surface and without further information regarding band state occupancy, no inference can be made to the change in the surface (quasi-) Fermi level. Furthermore, since here SPV is probed using a capacitance measurement, it fundamentally relates to changes in the net charge within the semiconductor system. A change in net charge may occur as a result of charge photogeneration and selective net extraction of one carrier or displacement of electronic carriers out of or into the system due to charge rearrangement within the sample. SPV hence resolves the relationship between electronic and ionic charge, the nature of which is sensitive to multiple material properties, including the recombination coefficients at the interfaces with both the atmosphere and the electrical contact.

We stress here that even if the number of photogenerated electrons and holes in the system increases upon illumination and results in greater quasi-Fermi level splitting, if there is no net change in the charge within the device system, the SPV signal will not change. Consequently, *the SPV signal arises from the change in electrostatic potential solely associated with changes in the net charge within the system*. Therefore effects such as localised photodoping [25], [26] would not give immediate rise to an SPV signal as device remains net charge neutral. Likewise, electronic or ionic rearrangement alone within the semiconductor system, while influencing the electrostatic potential, will not be sensed by a capacitance measurement.

To interpret our SPV results we modelled perovskite transport layer devices using coupled drift-diffusion equations. Driftfusion, as detailed in **SI Section 7**, solves the time-dependent continuity equations for electrons, holes, up to two ionic species, and Poisson's equation in one dimension within a semiconductor device stack[27], [28]. For the simulations in this work, we use a value for the Schottky defect density of $N_{SD}$ = $10^{18}$ cm$^3$, based on a range of calculated and experimentally measured values[29]–[39], outlined in detail in **SI Section 15**. Although not required for our SPV model, we assume that the perovskite shows close to intrinsic behaviour, which is generally a reasonably assumption in lead halide perovskites, with the possible exception of Pb/Sn systems which are quite p-type [21]. We only concern ourselves with ionic defects that have the ability to migrate, for example iodide vacancies resulting from Schottky defects, where a cation and anion are missing from their lattice sites [40]. We use current minimising potential boundary conditions (detailed in **SI Section 7**) to simulate the effect of the backing potential in the Kelvin probe experimental system.

The simulated device stack architecture was electrode/contact/perovskite/atmosphere (**Figure 3A**) such that the contact and perovskite layers are explicitly modelled, whereas the electrode and atmosphere regions are implicitly defined using boundary conditions. In this example, the Fermi energies of the perovskite and electron transport layer (ETL) are matched such that at equilibrium only a small amount of space charge is present at the contact and the potential profile is approximately uniform across the bilayer (see **SI Section 7**). The potential distribution, *V*, in the device depends on the spatial distribution of charge from electrons, holes, anions and cations, $\rho(x,t)$, and the boundary conditions, according to Gauss's Law:

$$V(x,t) = -\iint \frac{\rho(x,t)}{\varepsilon_0 \varepsilon_r(x)} dx dx,$$

*Equation 1*

where *x* is the spatial position, *t* is time, $\varepsilon_0$ is the permittivity of free space and $\varepsilon_r$ is the spatially dependent relative dielectric constant.

We reiterate that spatial rearrangement of the charge distribution alone will also not be detected by the Kelvin probe and hence not measured in the SPV signal. Only a net change in the total charge of the semiconductor system, will register as an SPV signal. This presents a barrier to directly reading out the SPV signal from a solution to a model using **Equation 1**, since the total change in electric potential across the device $\Delta V_\text{dev}$ is a combination of the measured SPV, $\Delta V_\text{SPV}$, and the change in potential from charge rearrangement alone $\Delta V_\text{red}$, such that on illumination:

$$\Delta V_\text{dev} = \Delta V_\text{SPV} + \Delta V_\text{red}.$$

The problem then is to find the change in the surface electrostatic potential solely associated with changes in electronic charge within the system, since we assume ionic charge carriers are confined to the perovskite layer. Our approach is to use the change in the areal charge density ΔQ (with units of C cm$^{-2}$) and the electronic component of the area-normalised capacitance $C_\text{el}$ (units: F cm$^{-2}$) due to electronic charge (see **SI Section 8** for details), such that

$$V_\text{SPV}(t) = \frac{\Delta Q(t)}{C_\text{el}}$$

*Equation 2*

ΔQ can be obtained by spatially integrating the change in volumetric space charge density Δρ directly following illumination of the system,

$$\Delta Q(t) = \int_0^d \Delta\rho(x,t)dx,$$

*Equation 3*

where $\Delta\rho(x,t) = \rho(x,t) - \rho_{t=0}(x)$. To verify the methodology and boundary conditions of the model, an SPV measurement was simulated on a single layer of n-type material which shows behaviour consistent with the traditional interpretation, and which shows excellent agreement with the SPV signal calculated using $\Delta Q$ and the appropriate capacitance relation (see **SI Section 8**).

Our model allows for SPV signals in excess of the open circuit voltage (and also with the opposite sign) and, for the mixed ionic-electronic conductors discussed here, the model suggests that the open circuit voltage and SPV are not directly relatable, as was previously proposed [23], [41], [42]. The SPV only corresponds to the component of the change in voltage associated with changes in overall charge within the semiconductor system. This distinction between the SPV signal and the actual change in electrostatic potential at the surface is critical to understanding these measurements and our apparently counter-intuitive SPV observations.

Our interpretation suggests that by modulating the perovskite back and front contact recombination coefficients, we may alter the quantity and sign of the net charge accumulated in the device. It has been shown that the concentration of trap states which mediate surface recombination can be reduced by passivation with oxygen [18]. In oxygen-containing ambient atmospheres, $O_2$ molecules influence the local geometry of iodine vacancies [43]. Oxygen acts to replace vacant iodine sites and restores the full octahedral coordination around Pb, which results in a well-passivated material. This interpretation is supported by previously reported simulated atomistic structure studies [44]. We now consider how the surface recombination influences the model system.

To parameterise that effect in the simulation, we alter electron and hole surface recombination velocities, $s_{n,r}$ and $s_{p,r}$, on the right hand side. In the air case (**Figure 2B)**, they have been set to zero for simplicity, representing a perfectly passivated surface (for more details of parameters, including electron and hole densities and results for a nearly perfectly passivated surface, see **SI Table 2** and **SI Section 14**). Simulation results for the unpassivated case (**Figure 2F)** have the same parameter set with the exception that $s_{n,r} = s_{p,r} = 10^4$ cm s$^{-1}$ for both carriers at the perovskite-atmosphere interface to represent the surface recombination induced by the oxygen free vacuum environment. Both simulated SPV measurements (**Figures 2B&F**) reproduce the characteristic response of experimental results, the approximate magnitude, and the atmosphere-dependent switch in polarity.

**Figures 2C&D** outline the mechanistic picture of the origins of SPV signal in the passivated perovskite which we now describe. Photogenerated electrons and holes are created at a constant rate *G* in the perovskite layer. The electrons are preferentially collected by the contact leaving a low concentration in the perovskite, both in the passivated and unpassivated cases. Electron-hole recombination in the system is dominated by processes at the interface, and surface in the un-passivated case (see **Figure 3**). This process occurs via delocalised traps or interfacial states described by Shockley Reed Hall recombination which is related to the surface or interface density of the electrons and holes, $n_s$ and $p_s$,

respectively. For a constant light intensity, the generation rate of electrons and holes (*G*) and recombination (*R*) rate will be equal, forming a dynamic equilibrium:

$$G = R \propto \frac{n_s p_s - n_i^2}{(p_s + p_t)/s_n + (n_s + n_t)/s_p}$$

*Equation 4*

where $n_i$ is the small intrinsic carrier concentration, $n_t$ and $p_t$ are constants (proportional to the density of traps) giving the density of trapped electrons and holes when their respective quasi-Fermi levels are at the trap energy (for simplicity we have assumed a single trap energy here). The recombination velocities $s_n$ and $s_p$ combined with the electronic charge and trap densities at the surfaces then control the rate of recombination. In the surface passivated case, the rate of recombination is mostly controlled by carrier concentrations at the perovskite/FTO interface, where the electron and hole populations across the interface overlap due to electrostatic attraction. At longer times (**Figure 1D**), mobile ions diffuse to the contact, replacing holes which diffuse away due to the difference in concentration gradients of the ion and hole populations. To maintain the dynamic equilibrium where the recombination rate must remain constant (according to **Equation 4**), we must have a net increase in the number of holes in the perovskite, and hence an increase in positive charge in the system resulting in a positive SPV signal. In the unpassivated case, the hole population does not accumulate in the bulk because of significant recombination with traps at the perovskite surface [18]. To maintain the dynamic equilibrium, the population of electrons in the contact must increase to avoid a drop in the recombination rate, resulting in a negative SPV signal. Thus, the behaviour of the SPV signal observed is dependent on processes at the buried interfaces and the physical interpretation cannot be limited to the top surface, as in highly doped semiconductors (in more detail, see **SI Section 6**).

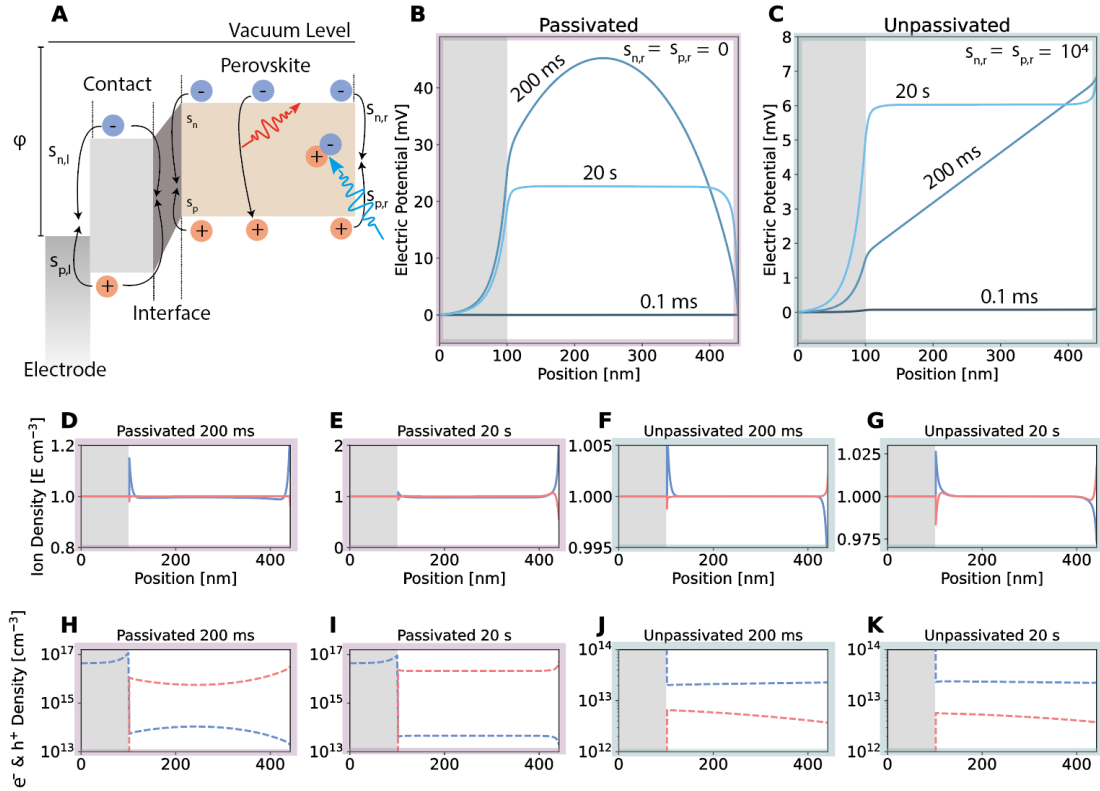

*Figure 3 A - Schematic of a simulated electrode/contact/perovskite/atmosphere device. The electron selective contact layer is attached to a floating potential electrode and the perovskite layer. Internal material interfaces are modelled as discrete interlayers (see Calado et al for a complete description[27]) and the thickness has been exaggerated for illustration. Electron, hole, cation and anion mobilities are defined in the text. The irradiance, is set throughout at 0.3 kW cm$^2$. **B & C** – Electric field in the passivated and unpassivated systems on an ETL, at 3 timepoints, 0.1ms, 200 ms and 20 s. **D-G** Ion migration in an perovskite bilayers on an ETL contact, the red and blue solid lines represent cations and anions, respectively. **H - K** – Electronic charge carrier populations, where the red and blue dashed lines represent holes and electrons, respectively.*

**Figure 3B** and **C** plot the electrostatic potential profiles for the passivated and unpassivated layers at different times throughout the SPV measurement. The second order derivative of the E-potential at the surface of the perovskite will determine the polarity of the SPV signal. Details on ionic and electronic charge in the passivated and unpassivated cases are given in **Figures 3D-K.** An SPV signal may also be recorded even without an idealised ETL or HTL transport layer as in **SI Figure 13**. In the case of an FTO contact, the FTO preferentially accepts electrons and generates a similar electric field throughout the system as an ETL would. This may be surprising as FTO is commonly thought of as a non-selective contact, however, the work function of FTO (since it is a degenerately doped n-type semiconductor) provides for a slight preference for electrons, which then induces a similar effect as to the ETL contact. This slight selectivity means it follows the fields built up in the ETL case. Additionally, for the FTO case, we also calculated the radiative recombination flux, which is consistent with experimental photoluminescence measurements (**SI Figure 4**).

We also simulated a hole transport layer (HTL) on the bottom surface (**SI Figure 14**) and find that consistent with reported measurements [22, Fig 3 & 5 ], the selectivity of the transport layer to either electrons or holes determines the polarity of the signal. The switch in polarity of the SPV signal between devices using HTL and ETL contact materials can be understood in terms of a switch in the sign of the space charge densities, both at the contact-perovskite interface and within the bulk of the perovskite, deriving from the contact selectivity to a single electronic carrier. However, not only the nature of the contact will determine the polarity. As we have shown, the recombination characteristics of the atmospheric interface will also alter the rate and direction of ion migration (more details in **SI Figure 14** for the migration of cations and ions in the perovskite bilayers for different contacts). Notably, it is possible to have cations or anions accumulate at the surface interface simply by modulating the passivation properties of the surface or the nature of the contact.

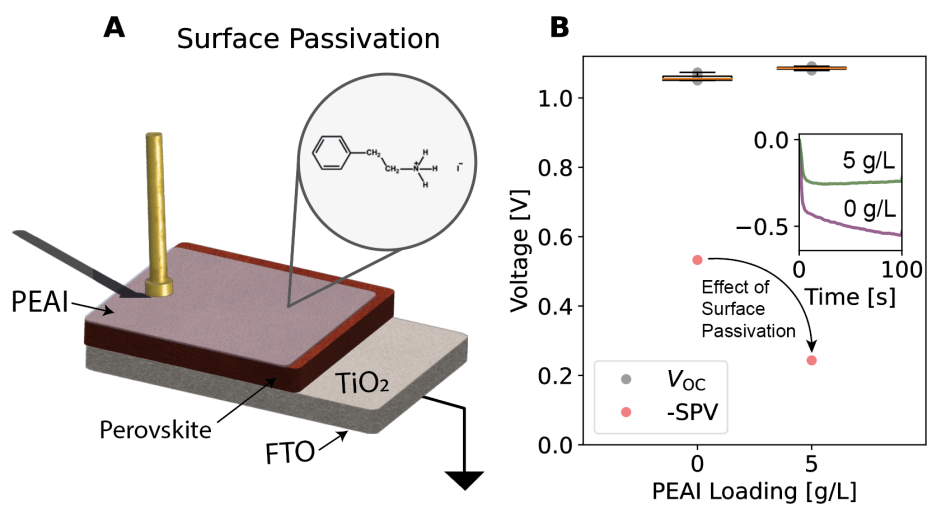

*Figure 4 A Schematic of a passivated arrangement -  FTO/TiO₂/Cs$_{0.05}$(FA$_{0.83}$MA$_{0.17}$)$_{0.95}$Pb(I$_{0.86}$Br$_{0.13}$)$_3$/PEAI. **B** In red, average value (for 10 seconds) of the SPV after 100 seconds illumination. In grey, open circuit voltage measurements of full solar cell stacks FTO/TiO₂/Cs$_{0.05}$(FA$_{0.83}$MA$_{0.17}$)$_{0.95}$Pb(I$_{0.86}$Br$_{0.13}$)$_3$/ PEAl/Spiro − OMeTAD/Au . Inset, SPV transient measurements of 5g/L (in green) and 0 g/L (in purple).*

Finally, to further validate our fndings and elucidate how the SPV response relates to device performance, we utilise chemical passivation on perovskite devices to establish a practical framework for predicting optimal device regimes. To this end, a perovskite device, in this case an FTO/TiO$_2$/$Cs_{0.05}(FA_{0.83}MA_{0.17})_{0.95}Pb(I_{0.86}Br_{0.13})_3$ arrangement (**Figure 4A**), was passivated with a layer of 2D cations, following an established passivation approach by application of a surface layer of phenylethylammonium iodide (PEAI) [45], [46] (see **Methods** for details). We evaluated the effectiveness of the passivation layer by monitoring SPV responses and open circuit voltage in an FTO/TiO$_2$/$Cs_{0.05}(FA_{0.83}MA_{0.17})_{0.95}Pb(I_{0.86}Br_{0.13})_3$/PEAI/Spiro-OMeTAD/Au perovskite device (see **Figure 4B** and see **SI Section 13** for details, as well as XRD). The transient SPV (**Figure 4B** inset) recorded in ambient conditions stabilised rapidly within 100 seconds upon PEAI loading of 5 g/L compared to the unpassivated device, and the SPV magnitude at 100 seconds reduced by 55% from 0.53V to 0.24V, whereas the open circuit voltage had a much smaller fractional change of 2.5%. This finding is consistent with our photoluminescence results, which suggest a reduction in non-radiative recombination rates (see **SI Section 13**). SPV hence provides a promising metric to track for optimizing device performance, while also reaffirming our conclusion that SPV does not directly measure open circuit voltage in a device lacking top contacts.

The origin of the biphasic response arises from the large differences in mobility between the electronic, cationic and anionic carriers. The rate of change of the signal observed for devices measured in vacuum can be directly correlated to a resistor-capacitor time constant for the redistribution of mobile ionic charge, $\tau \approx R_{ion} C_{ion}$ (see **SI Section 9**) [47]. However, measurement of the time constants given by any of the ionic capacitance relations can only be related to the quantity $N_{SD} \mu_{ion}^2$ (see details in **SI Section 9**). Given the large range of literature values for both $N_{SD}$ and $\mu_{ion}$ the direct relation of the time constant to defect density or ionic mobility is not immediately apparent (see **SI Section 15**). Nevertheless, under the simplistic assumption that the majority of mobile defects are of the Schottky type, the ratios of the time constants obtained from biexponential fitting gives the ratio of the cation and anion mobilities:

$$\frac{\tau_1}{\tau_2} = \frac{\mu_{cation}}{\mu_{anion}}$$

Repeated cycles of the light soaking treatment acts to reduce the resistor-capacitor time constant recorded in air (see **SI Figure 9A**). Since the ionic resistance is dependent on the quantity of mobile ionic defects, reductions in the rate may arise from the conversion of impurities such as lead to lead oxide or dissociation through pin-hole formation [18]. This physical mechanism may also explain why dark SPV signals on some perovskites devices quickly return to the initial value measured in the dark, whereas others do not [22], [48].

Conclusion

We have detailed that surface passivation recombination may be explored through SPV. The switch in the polarity of surface photovoltage along with a corresponding in the direction of the ionic distribution in the bilayer is dependent on the effectiveness of the surface passivation. This has important implications for the rational design of stability in devices since defect accumulation is a major pathway in degradation.

Importantly, we conclude that the SPV in mixed electronic ionic conductors can be understood in terms of the change in electric potential at the surface associated with changes in the net charge within the semiconductor system. The model presented here can account much of the phenomena seen in real-world measurements. The relationship electronic and ionic charge is highly sensitive to multiple material parameters including the selectivity of the contacts, ionic conductivity, and surface recombination properties. Surface photovoltage measurements (SPV) can reveal rich information on photo-excited dynamics of mixed electronic ionic conductors.

We have further experimentally verified avenues for the selective control of ion migration in perovskite bilayers, demonstrated here by modulating surface traps through surface passivation. We outline that relatively small changes in recombination rates at contacts can have drastic effects on the electric field and polarity of the device. The use of the SPV time constant, in the context of perovskite bilayers, may be used to identify relative changes in the density of ionic species in different samples, which could aid identification of species capable of reducing the formation of degradation sites, or improve electrical properties of memristors. More generally, our interpretation of SPV opens new avenues for resolving ion migration and charge transport which will aid understanding of mixed electronic ionic conductors.

## Methods

**Perovskite precursor solution and film deposition.**

Triple-cation perovskite $FA_{0.79}MA_{0.15}Cs_{0.06}Pb(I_{0.85}Br_{0.15})_3$ precursor solution was prepared by using $PbI_2 (1.1M)$, $PbBr_2 (0.22M)$, formamidinium iodide (1.00M), and methylammonium bromide (0.20M). These precursors were dissolved in a mixture of anhydrous DMF:DMSO (4:1 volume ratio, v:v) followed by addition of 5 vol% from CsI stock solution (1.5 M in DMSO). Lead compounds were purchased from TCI, the organic cations from Greatcell Solar, and cesium iodide from Alfa Aesar.

The glass substrates were cleaned via sonication in acetone and isopropyl alcohol for $10$ min in each solvent, followed by a treatment with oxygen plasma for $10$ min. The perovskite films were prepared by spincoating the perovskite solution on the glass substrates. We employed a two-step program in which the solution was spun at 1,000 and $4,000$ rpm for 10 and $30$ s respectively, and $110 \mu$ l of chlorobenzene was deposited on to the spinning substrate $10$ s prior to the end of the program. The samples were then annealed at $100°C$ for 1 hour. Synthesis and deposition of perovskite solutions were carried out in a dry nitrogen filled glove box with controlled relative humidity and oxygen ($H_2O$ level: $<$ 1ppm and $O_2$ level: $<$ 10ppm).

Fluorine doped tin oxide (FTO) Substrates were were cleaned in an ultrasonic bath in 2% (%v/v) Hellmanex III solution, deionized water, acetone, and isopropanol, consecutively. Between each solvent, substrates were washed and rinsed with the next solvent to minimize the residuals. Immediately before thin film deposition, 15 minutes of 100 W oxygen plasma or 20 minutes ultraviolet-ozone (UV-ozone) treatment was performed.

For mixed cation and anion perovskite (3D perovskite), two solutions (solution1 and solution2) were prepared and mixed as solution1:solution2 = 95:5 before use. Solution1 was composed of 1.5 M of formamidinium iodide (Greatcell Solar, 99.99%), 1.5 M of lead iodide (Tokyo Chemical Industry, 99.99%), 0.265 M of methylammonium bromide(Greatcell Solar, 99.99%) and 0.265 M of lead bromide (Sigma-Aldrich, >98%) in a mix solution of dimethylformamide (DMF)/dimethyl sulfoxide (DMSO) (4:1, %v/v). Solution2 was composed of 1.15 M of lead iodide (Sigma-Aldrich, 99%) and cesium iodide (Sigma-Aldrich, 99.9%) in a mix solution of DMF/DMSO (78:22, %v/v). The thin films of 3D perovskite were then deposited by spin-coating at 1000 rpm for 10 s and continuously at 4000 rpm for 30 s. At 5 s before the end, 110 µL of chlorobenzene was carefully dripped by micropipette at the center of the substrate. The samples were then annealed at 100 °C for 1 hour. For the PEAI or PEABr passivation, the samples were spun at 4000 rpm with a dynamic dripping of 5, 10 and 15 mg/mL PEAI or PEABr isopropanol solution at 5 s before the end.

For the electron transport layer (ETL), two solutions (ETL1 and ETL2) were prepared and mixed before use. ETL1 was composed of 1 mL of isopropanol (Sigma-Aldrich, anhydrous 99.5%) and 13.83 µL of 2M HCL(aq). ETL2 was composed of 1 mL of isopropanol and 145 µL of titanium isopropoxide (Sigma-Aldrich, 97%). Then, ETL1 was added into ETL2 slowly with vigorous stirring. The compact $TiO_2$ layer was deposited at 2000 rpm for 20 s and followed by 10 min annealing at 150 °C. After cooling down, samples were annealed in 500 °C for 30 min.

For the hole transport layer, 73 mg of Spiro-OMeTAD (Lumtech, 99%) was dissolved in 1 mL of chlorobenzene first, and then adding 28.8 µL of 4TBP (Sigma-Aldrich, 98%), 17.5 µL of LiTFSI (ACROS, 99%) acetonitrile solution (520 mg/mL) and 29 µL of FK209 (Sigma-Aldrich, 98%) acetonitrile solution (300 mg/mL) consecutively. The solution was always prepared no more than 24 hour before use to prevent degradation or forming clusters. A gold evaporated electrode (depth > 70nm) was applied to complete the device stack.

### Kelvin probe

Two Kelvin probe systems were used for this study. Relative humidity measurements were carried out on a RHC5050 (KP Technology) and vacuum measurement used a corner cube UHV020 (KP Technology) using 2 mm gold and stainless steel tips.

Surface photovoltage (SPV) measurements were taken using an APS04 (KP Technology). Test samples were grounded through the FTO in dark at least 30 min to reach equilibrium. Measurements were carried out with 20 s of dark, 100 s of illumination and 150 s of dark per cycle. SPV illumination was with a quartz tungsten halogen light source (Dolan–Jenner) at ~20 mW/cm2 intensity.

### JV Measurements

J-V scans were obtained using a solar simulator with AM1.5G filters (Oriel Instruments) and a Keithley 2400 Source Measure Unit. The solar simulator was warmed up for at least 30 min and the intensity was calibrated to 100 mW/cm2 with a silicon photodiode (Osram BPW21) prior to devices. Both reversed and forward scans were carried out with reverse scans first. The voltage was ranged from 1.6 to -0.4 V at a rate of 0.1 V/s.

### XRD

X-ray diffraction (XRD) was carried out using a Bruker D8 Discover with a Cu Ka (1.54 Å) source. Scans were done in a Bragg-Brentano set-up using a 0.3o divergent slit was applied to analyze perovskite crystal structure. The diffraction pattern is scanned over the angle 2θ from 5° to 45° with a step of 0.02983° with a scan time of 1 second per step. The method is used to identify PbI2 defect states with a characteristic peak at 12.67° and low dimensional perovskite (Ruddlesden-Popper phase).

### PL

Photoluminescence (PL) were carried out with FLS1000 photoluminescence spectrometer (Edinburgh Instruments). Samples were excited from the perovskite top layer with 405 nm incident light from a Xenon lamp. The emitted light was filtered by a 495 nm long-pass filter to eliminate interference from the incident light.

## Code Availability

All code used in this work is available from insert repository link in proof.

## Acknowledgements

T.K.B. gives thanks to the Centre for Doctoral Training in New and Sustainable Photovoltaics (EP/L01551X/2) and the NanoDTC (EP/L015978/1) for financial support. We acknowledge financial support by the EPSRC (EP/M006360/1) and the Winton Program for the Physics of Sustainability. This work received funding from the European Research Council under the European Union's Horizon 2020 research and innovation programme (HYPERION, grant agreement no. 756962). S.D.S. acknowledges support from the Royal Society and Tata Group (UF150033). The authors thank the EPSRC (EP/R023980/1) for funding. The Advance Imaging of Materials (AIM) facilities at Swansea University are supported by the European Regional Development Fund through the Welsh Government (80708) & EPSRC (EP/M028267/1). T.K.B., K.G. and S.D.S. thank Prof Iain Baikie for use of UHV Kelvin probes. Y.C.C acknowledges fincial support from Imperial President's PhD scholarship.

## Author Contributions

T.K.B. and K.G. performed the SPV and optical experiments for all samples other than PEAI study. Z.A-G prepared all samples other than the PEAI passivation study. Y-C.C., J.L., C.H., J.S.K developed the metholodgy and prepared the samples for the PEAI samples and the associated SPV and J/V measurements. T.D and J.M carried out the XRD measurements. P.C. and P.R.B.B developed Driftfusion. T.K.B and P.C. used Driftfusion in the context of simulating the SPV results. All authors contributed to discussions, analysis and writing of the manuscript.

## Competing Interests

S.D.S is a co-founder of Swift Solar, a company commercializing high-performance perovskite solar cells.

## Supplementary Information

The supplementary Information contains further information on SPV, UV/Vis and TCSPC experiments including controls and their replicates, an overview over the data analysis used for these methods, and more details on the simulation methods.